\documentclass[11pt,leqno]{article}

\usepackage{amsmath,amssymb,amsfonts,amsthm,mathrsfs,verbatim} 
\usepackage{algorithm}
\usepackage[noend]{algpseudocode}
\usepackage{setspace}
\usepackage{authblk}
\usepackage{caption}
\usepackage{geometry}
\usepackage{graphics} 
\usepackage{graphicx}  
\usepackage{hyperref}
\usepackage{makeidx}
\usepackage{color}
\usepackage{bm}

\theoremstyle{definition}

\newcommand{\R}{\mathbb{R}}
\newcommand{\E}{\mathbb{E}}

\def\Q{ \mathcal{Q} }
\def\w{\bm w}

\newcommand\blue[1]{\textcolor{blue}{#1}}

\title{A Framework for Treating Model Uncertainty in the Asset Liability Management Problem}

\date{(Published in JIMO: \blue{\url{10.3934/jimo.2023021}})}

\author[a,b]{Georgios I. Papayiannis}

\affil[a]{\footnotesize Section of Mathematics, Department of Naval Sciences, Hellenic Naval Academy, Piraeus, GR}
\affil[b]{\footnotesize Stochastic Modeling and Applications Laboratory, Athens University of Economics \& Business, Athens, GR}

\begin{document}
\maketitle

\begin{abstract}
The problem of asset liability management (ALM) is a classic problem of the financial mathematics and of great interest for the banking institutions and insurance companies. Several formulations of this problem under various model settings have been studied under the Mean-Variance (MV) principle perspective. In this paper, the ALM problem is revisited under the context of model uncertainty in the one-stage framework. In practice, uncertainty issues appear to several aspects of the problem, e.g. liability process characteristics, market conditions, inflation rates, inside information effects, etc. A framework relying on the notion of the Wasserstein barycenter is presented which is able to treat robustly this type of ambiguities by appropriate handling the various information sources (models) and appropriately reformu\-lating the relevant decisi\-on making problem. The proposed framework can be applied to a number of different model settings leading to the selection of investment portfolios that remain robust to the various uncertainties appearing in the market. The paper is concluded with a numerical experiment for a static version of the ALM problem, employing standard modelling approaches, illustrating the capabilities of the proposed method with very satisfactory results in retrieving the true optimal strategy even in high noise cases.
\end{abstract}


\section{Introduction and literature overview}

The asset liability management (ALM) problem is traditionally a standard topic of interest for financial institutions and insurance companies, and concerns the derivation of the optimal financial strategy for covering the potential liabilities at future times. This task is performed by appropriately selecting the investing portfolio, consisting of various financial assets and products (e.g. bonds, stocks, derivatives, etc), in order to counter the risk occurred by the potential liabilities. The relevant discussion on the matter initiated under the Markowitz's \cite{markowitz1952} mean-variance (MV) perspective in \cite{sharpe1990} and then discussed in the multi-stage framework \cite{leippold2004} and in the continuous-time setting \cite{chiu2006} under this principle. Subsequently, various modelling approaches concerning the problem's underlying dynamics and revisiting the relevant optimal control problem have been proposed, by considering more flexible and versatile models, e.g. stochastic interest-rate terms are considered in \cite{bian2021, pan2017}, jump-diffusion models are studied in \cite{chiu2014}, regime switching models are discussed in \cite{chen2008, shen2020, wei2013}, considerations for the inflation process are introduced in \cite{zhu2020}, more volatile models (beyond the typical geometric Brownian motion model) concerning the risky assets of the financial market are discussed in \cite{sun2020, zhang2016, zhang2022}, and many others. In all the aforementioned cases, the ALM problem is considered under the MV context, where a stochastic optimization problem is formulated and then treated through the typical optimal control techniques.

The uncertainty framework in the ALM problem has been so far considered in \cite{chen2022, yuan2021} where a reference probability model has been provided concerning the underlying dynamics of the problem and the decision maker although trusts the model, may decide to deviate at a level from it in order to derive a more robust financial strategy. Robustness property in modern decision making was formally introduced in a quantitatively framework about a decade ago in \cite{hansen2001robust, hansen2011robustness, maccheroni2006}. Clearly, the meaning of remaining robust to uncertainty when making a financial decision, refers to the low degree of sensitivity of the deduced optimal decision with respect to the scenario that materializes. In this paper, a more complex decision making framework is considered, under which model ambiguity appears to various aspects of the ALM problem and there is available a multitude of plausible models (scenarios) that could describe the future states of the market with unknown validity. This is a much more realistic framework especially in periods where the market enters unstable states due to e.g. pandemic effects, economy instabilities caused by wars, energy crisis, etc. Under such a working framework, the robustness of the decision making process is twofold and refers to (a) the optimal action/decision/strategy to be chosen, and (b) aggregation/combination of the provided information (models) on which strongly depends the optimal strategy to be selected.

In this work, a framework that is able to handle model uncertainty issues that may appear in financial markets affecting significantly the ALM problem is introdu\-ced. The designing of effective and robust to model uncertainty portfolio selection rules are discussed to cover the potential liabilities of the interested financial institu\-tions or insurance companies. The proposed approach is based on the aggregation of the models provided by various information sources, identified by certain probability models, through barycentric approaches that have been recently introduced in decision theory \cite{petracou2022decision} and in financial risk management \cite{papayiannis2018convex}. The presented approach is directly implemented in designing the decision rule for selecting the optimal strategy (investing portfolio), that will remain robust to model uncertainty, reducing the exposure of a position to financial risk. Moreover, a numerical experiment is performed, employing standard models that are used in ALM practice, to illustrate the capabilities of the proposed methodology and testing its sensitivity with respect to different levels of information homogeneity.

\section{The asset liability management problem}\label{sec-2}

Let $(\Omega, \mathcal{F},\mathcal{Q})$ be a complete probability space on which an $n$-dimensional standard Brownian motion ${\bm W}(\cdot) = ( W_1(\cdot), ..., W_n(\cdot))^T$ and an $m$-dimensional Brownian motion ${\bm B}(\cdot) = (B_1(\cdot),..., B_m(\cdot))^T$ are defined where $\mathbb{F} := \{\mathcal{F}_t\}_{t \in [0,T]}$ denotes the augmented filtration generated by $({\bm W}(\cdot), {\bm B}(\cdot))$ with $T>0$ being a fixed future time. Note that ${\bm W}(\cdot)$ represents the part of stochasticity and risks introduced by the financial market, while ${\bm B}(\cdot)$ represents other sources of risk faced by the financial institution or the  insurance company (e.g. claim risks for insurance companies), respectively. On the setting introduced above, $\mathcal{Q}$ is a real-world probability measure concerning the possible states of the world represented in the set $\Omega$ and $\mathcal{F}_t$ denotes the information that is available up to time $t$. We typically assume that the market is free of arbitrage, frictionless and complete (however the latter assumption could be dropped in order to provide a more general framework).

Assume that the financial market consists of a risk-free asset (a bond) and $n$ risky assets. The bond's dynamics are described by the model
\begin{eqnarray}
	dS_0(t) = r(t) S_0(t) dt, \,\,\, S_0(0) = s_0 >0,
\end{eqnarray}
where $r(\cdot)$ denotes the instantaneous interest rate in which stochasticity could be introduced considering an Ornstein-Uhlenbeck (OU) process of the form
\begin{eqnarray}\label{r-dyn}
	dr(t) = \kappa (R_0 - r(t))dt + {\bm \sigma}_r^T d{\bm W}(t), \,\,\, r(0)=r_0>0
\end{eqnarray}
with $R_0>0$ denoting the long-term average interest-rate value, parameter $\kappa>0$ introducing the relevant mean-reversion effect and ${\bm \sigma}_r := (\sigma_{r,1}, ..., \sigma_{r,n})^T \in \R_+^n$ denoting the volatility vector that explains the process variability with respect to the financial market factors. The typical model that is employed for describing the risky asset's dynamics is the geometric Brownian motion (GBM) model. In particular, for the $i$-th risky asset the stock price dynamics are modelled as
\begin{equation}\label{S-dyn}
	dS_i(t) = S_i(t) \left[ \mu_i dt + {\bm \sigma_i}^T d{\bm W}(t) \right], \,\,\, S_i(0) = s_i
\end{equation}
with $\mu_i$ denoting the relevant drift term to the $i$-th risky asset and the vector ${\bm \sigma}_i := (\sigma_{i,1},...,\sigma_{i,n})^T \in \R_+^{n}$ denoting the volatility coefficients. In a more compact form we may represent the dynamics for all risky assets $\widetilde{S}(t):=(S_1(t),...,S_n(t))^T$ by the model
\begin{equation}
	d\widetilde{S}(t) = diag(\widetilde{S}(t))\left[ {\bm \mu} dt + {\bm \sigma} d{\bm W}(t)\right], \,\,\, \widetilde{S}(0) = {\bm s}_0 \in \R^n,
\end{equation}
where ${\bm \mu} := (\mu_1, ..., \mu_n)^T\in \R^{n}$, ${\bm \sigma} := [ {\bm \sigma}_1, ..., {\bm \sigma}_n]^T \in \R_+^{n\times n}$ and ${\bm s}_0 := (s_1, ..., s_n)^T\in\R^n$. Note that in this setting, ${\bm W}(\cdot)$ represents the total set of stochastic factors affecting the financial market (both the bond and stock prices). However, stock prices and bond prices may depend on different groups of factors and this could be specified by setting accordingly the respective volatility vectors. Assume further that the financial institution/insurance company considers a Brownian Motion (BM) model with drift as an aggregate model for its liability process (or in other terms of its cumulative claim process) $L(\cdot)$ of the form
\begin{eqnarray}\label{L-dyn}
	dL(t) = \alpha dt + {\bm \beta}^{T} d{\bm W}(t) + {\bm \gamma}^{T} d{\bm B}(t), \,\,\, L(0)=\ell_0 >0
\end{eqnarray}
where $\alpha$ denotes the drift (average claim rate) for the aggregate liability process, ${\bm \beta} := (\beta_1, ..., \beta_n)^T \in \R^n$ the volatility coefficients related to the financial market and ${\bm \gamma} := (\gamma_1,...,\gamma_m)^T \in \R^m$ the volatilies related to other sources of risk for the insurance activities. This model approximates the typical Cr\'amer-Lundberg model that is used for representing the claim process. The considered model (Brownian motion with drift $\alpha$) assumes that $\alpha t$ is the mean liability that may occur at time instant $t$ while the stochastic terms model the possible fluctuations around this mean (see e.g. \cite{baltas2012} and references therein). However, note that some care should be taken in tuning the volatility parameters to avoid negative values for the aggregate liability.

The construction of an investment portfolio $X(t)$ to cover the liabilities that will occur at the future time $t=T$ for the insurance company is considered here in a static setting. That means that at time $t=0$ needs to be determined how a certain amount of wealth, $x_0 \in \R$, should be optimally allocated by the risk management desk to the available assets in order to minimize the effect of future short positions due to the claims that might occur. The static setting is in many cases much preferable than the continuous allocation of wealth amounts to the various available assets (especially in relative short time periods) to avoid extra transaction costs or due to other reasons, e.g. regulations, market imperfections, liquidity constraints, etc. In this perspective, assume that $\theta_i$ denotes the proportion of wealth to be invested to the $i$-th security for $i=0,1,...,n$. In that case, the investment portfolio value at time $t=T$ is
\begin{eqnarray}
X^{\bm \theta}(T) = \theta_0 S_0(T) + \sum_{i=1}^n \theta_i S_i(T) = {\bm \theta}^T {\bm S}(T)
\end{eqnarray}
where ${\bm S}(t) := (S_0(T), S_1(T), ..., S_n(T))^T$ and ${\bm \theta} := (\theta_0, \theta_1, ..., \theta_n)^{T} \in \R^{n+1}$. The set $\Theta \subset \R^{n+1}$ of admissible investing strategies should be determined by the company's constraints and preferences concerning financial decisions like liquidity, maximum and minimum amount to be allocated to certain type of assets, etc. The wealth process (investing portfolio value) $X(\cdot)$ is a controllable process depending on the interest rate and stock prices dynamics determined in \eqref{r-dyn} and \eqref{S-dyn} and to clarify the dependence on ${\bm \theta}(\cdot)$ in what follows the notation $X^{\bm \theta}(\cdot)$ is adopted, while the liability is considered as an uncontrollable BM process with drift as defined in \eqref{L-dyn}. In this framework, the problem is treated as an one-stage optimization problem since the strategy ${\bm \theta}$ to be derived refers only to the optimal decision to be made at time $t=0$ and it will not change until the horizon $T$. According to the MV principle, the static version of the asset liability management problem considered here is represented by the mean variance optimization problem
\begin{eqnarray}\label{mv-2}
	&&\min_{{\bm \theta} \in \R^n} \E_{\mathcal{Q}}\left[  (X^{\theta}(T) - L(T) - \zeta )^2 \right] \nonumber \\
	&&\mbox{subject to} \nonumber \\
	&&\left\{
	\begin{array}{l}
		\E_{\mathcal{Q}}[ X^{\theta}(T) - L(T) ] \geq \zeta\\
		(L(\cdot), X^{\bm \theta}(\cdot)) \mbox{ obtained  by the dynamics } \eqref{r-dyn}, \eqref{S-dyn} \mbox{ and } \eqref{L-dyn}\\
		{\bm \theta} \in \Theta.
	\end{array}\right.
\end{eqnarray}
where $\zeta \in \R$ parameterizes the acceptable mean levels of the surplus process $X^{\bm \theta}(\cdot) - L(\cdot)$ at time $T$, $x_0 \in \R$ determines the total wealth to be invested, $\Theta \subset \R^n$ represents the set of all acceptable wealth amount allocations (possibly containing constraints for the lower and upper wealth amount levels to be invested to each asset and others) and $\E_{\mathcal{Q}}[\cdot]$ denotes the expectation with respect to the probability measure $\mathcal{Q}$. For the sake of simplicity denote $X_T := {\bm \theta}^{T} {\bm S}_T$ where ${\bm S}_T := (S_0(T), S_1(T),...,S_n(T))^T$. Then, deriving the associated Langrangian to the problem the optimal investing portfolio in terms of the wealth amount to be invested to each one of the available assets (both bond and stocks) is
\begin{eqnarray}\label{opt-port}
	{\bm \theta}_*(\lambda) = \widetilde{C} [C_{{\bm S} L} + (\zeta + \lambda) {\bm m}_{S}] + \frac{ C_{\bm S}^{-1} {\bm 1} x_0 }{{\bm 1}^{T} C_{\bm S}^{-1} {\bm 1}}
\end{eqnarray}
where ${\bm m}_{\bm S} := \E_{\Q}[{\bm S}_T]$, ${\bm 1}=(1,1,...,1)^T \in \R^{n+1}$, $\widetilde{C}:= C^{-1}_{\bm S} - C^{-1}_{\bm S} {\bm 1} {\bm 1}^T C^{-1}_{\bm S} / ( {\bm 1}^T C_{\bm S}^{-1} {\bm 1} )$, $C_{\bm S} := \E_{\Q}[{\bm S_T}{\bm S}_T^{T}]$, $C_{{\bm S} L} := \E_{\Q}[{\bm S}_T L_T]$ and $\lambda \in \R_+$ denotes the associated Lagrangian multiplier to the inequality constraint $\E_{\Q}[X_T-L_T]\geq \zeta$ for any specified level $\zeta \in \R$. Depending on the problem data and parameters we distinguish between the following two cases:
\begin{enumerate}
	\item {\it Non-active constraint case:} optimal investing portfolio is the portfolio expressed in \eqref{opt-port} by substituting $\lambda_*=0$.
	\item {\it Active constraint case:} optimal investing portfolio is the portfolio expressed in \eqref{opt-port} by substituting
	\begin{eqnarray}
		\lambda_* = \frac{1}{ {\bm m}_{\bm S}^T {\bm m}_{\bm S} } \left[ m_{L} + \zeta(1-{\bm m}_{\bm S}^T {\bm m}_{\bm S}) - {\bm m}_{\bm S}^T (\widetilde{C}C_{{\bm S} L} + C^{-1}_{\bm S} {\bm 1} x_0)\right]
	\end{eqnarray}
	where $m_L := \E_{\Q}[L_T]$ for any determined value of $\zeta \in \R$.
\end{enumerate}

\section{The model uncertainty framework}\label{sec-3}

In financial mathematics, even if a closed-form solution can be obtained concerning the optimal decision (like the static ALM problem discussed in this work \eqref{opt-port}), its calculation relies on the assumption that a probability model $\Q$ describing the stochasticity related to the problem under study is known. It is usual in practice, the decision maker to face the problem of model uncertainty, i.e. several scenarios could happen in the near future but it is extremely difficult, if not impossible, to distinguish which scenario will actually occur. In such a case, the various possible models describing the forthcoming situation must be treated in a robust manner, to reduce the sensitivity of the financial decision to be made with respect to the scenario that actually materializes. This task is twofold since the decision maker should: (a) appropriately combine the received information to a model that can be used for the discounting of the decision, and (b) the resulting aggregate model to be as robust as it gets to the contaminated noise of the prior set of models. A framework that is appropriate for handling model uncertainty in problems of this type is proposed in \cite{papayiannis2018convex} and \cite{petracou2022decision} under the perspectives of financial risk quantification and group decision making, respectively. In both approaches, various scenarios or opinions are identified by certain probability models, and the notion of the Wasserstein barycenter (roughly the sense of median in the space of probability models) is employed to estimate an aggregate model that represents/condenses the received information and which is then used to derive a robust decision rule. In what follows, these ideas are carefully employed under the context of the asset liability management under model ambiguity however the proposed framework is directly applicable to numerous others problems in finance.

Assume that the insurance company which desires to derive its investing portfolio, faces various kinds of uncertainties concerning the stochastic processes that constitu\-te the ALM problem. Due to the lack of sufficient statistical data or due to inability to observe the related random factors, uncertainty issues appear concerning: (a) the appropriate model to describe the surplus process and its components (wealth process, interest rate process, risky assets models, etc) either in terms of the exact parameters or in terms of the kind of the statistical model to be used, (b) significant financial market shifts which cannot be predicted by the collected data but only from sources of inside information, (c) uncertainties related to the insurance company's activities to new sectors where no previous experience exists. The risk manager in charge, to counter the uncertainty concerning the true (but unknown) model $\mathcal{Q}$, is provided with a collection of $N$ models $\mathfrak{Q} = \{\mathcal{Q}_1, \mathcal{Q}_2, ..., \mathcal{Q}_N\}$ by her/his network of market agents/sources. Each model possibly contains different fragments of reality and therefore trust cannot be fully allocated to just one of these models. The major issue arising here, is how to efficiently aggregate the available information into a single model which will be used to robustly discount the outcomes of the financial decisions to be made.

A robust and efficient approach in combining different beliefs into a single aggregate model has been proposed in \cite{papayiannis2018convex} and \cite{petracou2022decision} where the notion of the Wasserstein barycenter (derived from the generalized mean sense offered by the notion of Fr\'echet mean \cite{frechet1948elements}) is combined within the frameworks of convex risk measures and expected utility to properly discount decisions under model ambiguity. The Wasserstein distance is a proper metric in the space of probability models (see for technical details \cite{santambrogio2015optimal, villani2021topics}) and between two probability models $\Q_1, \Q_2 \in \mathcal{P}(\Omega)$\footnote{where $\mathcal{P}(\Omega)$ denotes the space of probability models with support in the set $\Omega$} is computed (according to Kantorovich formulation) as the minimal cost of the problem
\begin{equation}
	W_2^2(\Q_1, \Q_2) = \min_{\Gamma \in \Pi(\Q_1, \Q_2)} \int_{\Omega \times \Omega} \|x-y\|_2^2 d\Gamma(x,y)
\end{equation}
where $\Pi(\Q_1, \Q_2)$ is the set of transport plans with marginals the measures $\Q_1, \Q_2 \in \mathcal{P}(\Omega)$. Since the Wasserstein distance is fully compatible with the geometry of the space of probability models, constitutes a very effective instrument for the quantification of discrepancies between various opinions which can be identified by certain probabi\-lity models. On the examined framework where a prior set with a multitude of models is provided, the notion of Wasserstein barycenter is employed (see e.g. \cite{agueh2011barycenters} for technical discussion), i.e. an appropriate analog of the mean sense in the space of probability models (opinions), to derive a single aggregate model comprising all the available information from the set $\mathfrak{Q}$. Given a weighting vector $\w \in \Delta^{N-1}$\footnote{denoting the unit simplex, i.e. $\Delta^{N-1} := \{ \w\in\R^N\,\, : \,\,\sum_{i=1}^N w_i = 1, \,\,\, w_i \geq 0, \,\,\, \forall i=1,2,...,N\}$}, which can be considered as each model's contribution to the final aggregate, the Wasserstein barycenter is defined as the minimizer of the respective Fr\'echet variance
\begin{equation}\label{wbar}
	\Q_*(\w) = \arg\min_{\Q \in \mathcal{P}(\Omega)} \sum_{i=1}^N w_i W_2^2(\Q,\Q_i).
\end{equation}

In general, problem \eqref{wbar} does not admit a closed-form solution, however for the case that all models in $\mathfrak{Q}$ are members of a Location-Scatter family (e.g. Gaussian), which is the case in the discussed setting with GBM and/or Ornstein-Uhlenbeck type models, the barycenter of the set $\mathfrak{Q}$ can be characterized through its parameters in a semi-closed form. In particular, the location of $\Q_*(w)$ is represented as the weighting average
\begin{equation}\label{wloc}
	m_B(\w) = \sum_{i=1}^N w_i m_i
\end{equation}
with $m_i$, $i=1,2,...,N$ denoting the location parameters of models $\Q_1,\Q_2,...,\Q_N$, while the dispersion characteristics of $\Q_*(\w)$ are represented by the covariance matrix $C_B$ which satisfies the matrix equation
\begin{equation}\label{wcov}
	C_B = \sum_{i=1}^N w_i \left( C^{1/2}_B C_i C^{1/2}_B \right)^{1/2}
\end{equation}
with $C_i$ for $i=1,2,...,N$ denoting the covariance matrices of models $\Q_1,\Q_2,...,\Q_N$. The matrix $C_B$  can be obtained numerically by the fixed-point scheme  proposed in \cite{alvarez2016}. For the case of non Location-Scatter models, numerical schemes based on the entropic regularization of the main problem to achieve higher convergence rate have been proposed in the literature (see e.g. \cite{carlier2017convergence, clason2021entropic, cuturi2013sikhorn}).

The barycenter stated in \eqref{wbar} serves as the aggregate model of the set $\mathfrak{Q}$ depending on the weighting vector $\w$. The latter can be realized as a sensitivity parameter chosen by the risk manager, determining the level of contribution/infuence of each provided model in $\mathfrak{Q}$ to the model that will be actually used to discount the desicions. For instance, if $w_i=0$ for a certain $i$ is set, then this particular model is omitted as unrealistic or untrustworthy. Similarly, if $w_i=1$ is chosen, then the decision maker fully allocates her/his interest to this particular model and omits the rest. Equal weighting, i.e. $w_1=w_2=...=w_N=1/N$, corresponds to the typical barycenter case where no extra information about the credibility of each model is available. In the case that these models concern different scenarios that could be realilzed (and they do not represent just different opinions), the choice of the discounting probability measure in this way allows for determining a robust strategy across scenarios avoiding to concentrate to just one of them. In this perspective, $\w$ could be understood as the probability (possibly in a subjective view) for each one of the scenarios to occur. Weighting the various scenarios and not concentrating to just one of them, should lead to optimal decisions/strategies which financial output will be less affected by the scenario that actually occurs. This is the essense of robustness, however the choice of weighting vector $\w$ and the quality of the information provided by the prior set $\mathfrak{Q}$ (does it contains the ``true'' scenario?) can act as sensitivity parameters and should be carefully chosen and assessed. However, hybrid schemes combining aversion preferences with data-driven approaches in updating the weighting vector whereas new information batches are available can be developed by applying appropriate scoring rules (see, e.g. \cite{papayiannis2018learning, papayiannis2018model}).

Clearly, there is always the chance that no model in $\mathfrak{Q}$ is realized or the investor does not fully trust none of these models. In that case, the investor may quantify her/his aversion variationally in the spirit of \cite{hansen2001robust, maccheroni2006} and the deduced model would be a distorted version of the barycenter model as computed by the set $\mathfrak{Q}$ (please see \cite{papayiannis2018convex} and the related results therein).  The level of distortion depends on the level of reliability that the investor allocates to the provided information with the special cases corresponding to the barycenter model (fully trust in $\mathfrak{Q}$) and the most distorted model (no trust in $\mathfrak{Q}$) which corresponds to the deep uncertainty case. In this work, only the case where the decision maker is fully confident regarding the plausibility of the provided models is discussed.

\section{Robust formulation of the ALM problem}\label{sec-4}

Following the discussion in Section \ref{sec-3}, the ALM optimization problem stated in \eqref{mv-2} is revisited. It is clear that even if the models used for representing the dynamics of the involved processes $X(\cdot)$ and $L(\cdot)$ are widely acceptable, different perspectives (scenarios or opinions) concerning the related parameters could lead to much different derivations for the probability measure $\Q$ under which the investing decision should be discounted. This probability model contains information concerning the dynamics under which the processes $X(\cdot), L(\cdot)$ evolve and of course their relations to the stochastic factors included in ${\bm W}(\cdot)$ (financial market factors) and ${\bm B}(\cdot)$ (insurance market factors). The collected opinions/estimations will form the prior set $\mathfrak{Q}$ consisting of the derived probability models which depending the market situation and the states of economy could be quite close or even conflicting. An appropriate choice of model should be robust to these uncertainties, i.e. a robust model should lead to a decision that whatever scenario materializes, the optimal derived decision should be scenario-insensitive. Clearly, the term robustness in this work, refers to the sensitivity of the optimal decision to the probability model under which the financial decision is to be discounted, to possible inefficiencies concerning the various characteristics of the stochastic processes that contribute to the surplus amount output. In what follows, the prior set is assumed that provides sufficient information to retrieve the true situation, i.e. the true probability model (unknown) should be able to be approximated by the models provided by the prior set.

From the formulation of problem \eqref{mv-2} although is stated statically with respect to a certain time horizon $T$, it is evident that the possible outcomes of the surplus process variance depend on the probability model describing the dynamics of the related processes. Since, there is uncertainty regarding the true model $\Q$, and given the provided multitude of models $\mathfrak{Q}$, a robustified version of the problem \eqref{mv-2} is considered with respect to the discounting probability model. In particular, problem \eqref{mv-2} can be represented in a robust formulation as the minimax problem
\begin{eqnarray}\label{mv-3}
	&&\min_{{\bm \theta} } \max_{\Q \in \mathfrak{Q}_{\epsilon}} \E_{\mathcal{Q}}\left[  (X^{\theta}_T - L_T - \zeta )^2 \right] \nonumber \\
	&&\mbox{subject to} \nonumber \\
	&&\left\{
	\begin{array}{l}
		\E_{\mathcal{Q}}[ X^{\theta}_T - L_T ] \geq \zeta \\
		(L(\cdot), X^{\bm \theta}(\cdot)) \mbox{ obtained  by the dynamics } \eqref{r-dyn}, \eqref{S-dyn} \mbox{ and } \eqref{L-dyn}\\
		{\bm \theta} \in \Theta
	\end{array}\right.
\end{eqnarray}
where $\mathfrak{Q}_{\epsilon}$ denotes the set of plausible probability models according to the aversion preferences of the investor from the set of the provided models $\mathfrak{Q}$. The set $\mathfrak{Q}_{\epsilon}$ can be expressed as (using the Wasserstein metric sense)
\begin{equation}
	\mathfrak{Q}_{\epsilon} := \left\{ \Q \in \mathcal{P}(\Omega) \,\, : \,\, \sum_{i=1}^N w_i W_2^2(\Q,\Q_i) \leq \epsilon, \,\, \forall \w\in\Delta^{N-1} \right\}
\end{equation}
where $\epsilon>0$ denotes the sensitivity parameter quantifying the aversion intensity and $\w$ the weighting vector for the models in prior set $\mathfrak{Q}$. Setting the minimal value $\epsilon_*$ that can be attained for the Fr\'echet variance of a model $\Q$ from the set $\mathfrak{Q}$, is equivalent to employ the Wasserstein barycenter of the set (as defined in \eqref{wbar}) as the discounting measure. Clearly, taking values $\epsilon > \epsilon_*$ will lead to deformations of the barycentric model providing even worse estimates for expected loss compared to the ones that can be derived by trusting only the information provided in $\mathfrak{Q}$. Problem \eqref{mv-3} essentially means that the investor chooses the investing portfolio so as to minimize the maximum surplus process variance around the target value $\zeta$ over the set of possible probability laws for the surplus value at $T$, thus making an investing decision under the worst-case scenario (since the robust version of the problem is expressed as a minimax problem of the variance with respect to the admissible set of probability models). This problem can also be realized as a game where nature plays against the decision maker (insurance company), where the first player (nature) chooses the model that will produce the worst outcomes (losses) for the other side, while the second player (insurance company) seeks to be protected against worst-case outcomes.

In the discussed setting, the liability process as defined in \eqref{L-dyn} is described by a BM model with drift, the interest-rate process dynamics are captured by an OU model as determined in \eqref{r-dyn}, while the stock prices dynamics determined by the typical GBM model defined in \eqref{S-dyn}. It is well known, that for all the above stochastic processes the following closed-form solutions are obtained for any $t\in(0,T]$
\begin{eqnarray}
	\begin{array}{l}
	L(t) = \ell_0 + \alpha t + \sqrt{t} \left( \int_0^t  {\bm \beta}^T d{\bm W}(s) +  \int_0^t {\bm \gamma}^T d{\bm B}(s) \right)  \\
	r(t) = R_0 + e^{-\kappa t} (r_0 - R_0)  + \sqrt{\frac{1-e^{2 \kappa t}}{2\kappa}} \int_0^t {\bm \sigma}_r^{T} d{\bm W}(s)\\
	\log S_i(t) = \log s_{i} + \mu_i t + \sqrt{t} \int_0^t {\bm \sigma}_i^T {\bm W}(s), \,\,\,\, i=1,2,...,n.
	\end{array}
\end{eqnarray}

However, for any fixed time $t \in (0,T]$ the above closed-form solutions can be represented by the random variables
\begin{eqnarray}\label{model-rvs}
	\begin{array}{l}
	L_t = \ell_0 + \alpha t + \sqrt{t} \left( {\bm \beta}^T {\bm Z}_W + {\bm \gamma}^T {\bm Z}_B \right)  \\
	r_t = R_0 + e^{-\kappa t} (r_0 - R_0)  + \sqrt{\frac{1-e^{2 \kappa t}}{2\kappa}} {\bm \sigma}_r^{T} {\bm Z}_W\\
	\log S_{i,t} = \log s_{i} + \mu_i t + \sqrt{t} {\bm \sigma}_i^T {\bm Z}_B, \,\,\,\, i=1,2,...,n
	\end{array}
\end{eqnarray}
where the random variables ${\bm Z}_{W}:\Omega \to \R^n$ and ${\bm Z}_B: \Omega \to \R^m$ follow Normal distributions with zero locations and covariance (correlation) matrices ${\bm C_W}$ and ${\bm C_B}$, respectively. Since these random variables provide instantaneous information derived from the Brownian processes ${\bm W}(\cdot)$ and ${\bm B}(\cdot)$, and given that these informati\-on sources are considered by assumption as independent (i.e. $\E_{\mathcal{Q}}[dW_i(t) dB_j(t)] = 0$ for any $i,j$) then both parts of information (and their dependence) can be represented by the random variable
\begin{eqnarray}\label{Z-prob}
	{\bm Z} = \begin{pmatrix}{\bm Z}_{W}\\ {\bm Z}_{B} \end{pmatrix} \sim {\bm N}_{n+m}\left( {\bm m}_Z,
	{\bm C}_Z   \right)
\end{eqnarray}
with
\begin{eqnarray}
	{\bm m}_Z := \begin{pmatrix}{\bm 0}_n\\ {\bm 0}_m \end{pmatrix} \in \R^{n+m}, \,\,\,\, {\bm C}_Z := \begin{pmatrix}
		{\bm C}_W & {\bm 0}_{n\times m}\\
		{\bm 0}_{m \times n} & {\bm C}_B
	\end{pmatrix} \in \R^{(n+m)\times(n+m)}.
\end{eqnarray}

In fact, the random variable defined in \eqref{Z-prob} allows for a more convenient modelling of the random behaviour concerning the stock prices and the liability process at any fixed time instant $t$. Specifically, for any $t$ the $(n+2)$-dimensional Gaussian random variable
\begin{eqnarray}\label{Y-prob}
	{\bm Y}_t := ( L_t, r_t, \log S_{1,t}, \ldots, \log S_{n,t})^{T} \sim {\bm N}_{n+2} \left( {\bm m}_t, {\bm C}_t \right)
\end{eqnarray}
is characterized by the location and dispersion parameters
\begin{eqnarray}
	&&{\bm m}_t := \begin{pmatrix}
		\ell_0 + \alpha t \\
		R_0 + e^{-\kappa t}(r_0 - R_0) \\
		\log s_{1} + \mu_1 t\\
		\vdots \\
		\log s_{n} + \mu_n t
	\end{pmatrix}, \\
&&{\bm C}_{t} :=
\begin{pmatrix}
	t {\bm b}^T {\bm C}_Z {\bm b} & c(t) {\bm b}^T {\bm C}_Z \widetilde{\bm \sigma}_r & t {\bm b}^T {\bm C}_Z \widetilde{\bm \sigma}_1 & \cdots & t {\bm b}^T {\bm C}_Z \widetilde{\bm \sigma}_n \\
	
	c(t) \widetilde{\bm \sigma}_r^T {\bm C}_Z {\bm b} &
	\tilde{c}(t) \widetilde{\bm \sigma}_r^T {\bm C}_Z \widetilde{\bm \sigma}_r &
	c(t)\widetilde{\bm \sigma}_r^T {\bm C}_Z \widetilde{\bm \sigma}_1 &
	\cdots & c(t)\widetilde{\bm \sigma}_r^T {\bm C}_Z \widetilde{\bm \sigma}_n  \\
	
		t \widetilde{\bm \sigma}_1^T {\bm C}_Z {\bm b} & c(t) \widetilde{\bm \sigma}_1^T {\bm C}_Z \widetilde{\bm \sigma}_r & t \widetilde{\bm \sigma}_1^T {\bm C}_Z \widetilde{\bm \sigma}_1 & \cdots &t \widetilde{\bm \sigma}_1^T {\bm C}_Z \widetilde{\bm \sigma}_n \\
	
	\vdots & \vdots & \vdots &\ddots  & \vdots\\
	
	t \widetilde{\bm \sigma}_n^T {\bm C}_Z {\bm b} & c(t) \widetilde{\bm \sigma}_n^T {\bm C}_Z \widetilde{\bm \sigma}_r & t \widetilde{\bm \sigma}_n^T {\bm C}_Z \widetilde{\bm \sigma}_1 & \cdots &t \widetilde{\bm \sigma}_n^T {\bm C}_Z \widetilde{\bm \sigma}_n
\end{pmatrix} \nonumber
\end{eqnarray}
where $\widetilde{\bm \sigma}_r := ( {\bm \sigma}_r^T, {\bm 0}_{m}^T)^T$, $\widetilde{\bm \sigma}_i := ( {\bm \sigma}_i^T, {\bm 0}_{m}^T)^T$ for all $i=1,2,...,n$, ${\bm b} = ({\bm \beta}^T, {\bm \gamma}^T )^T$, $c(t) = \sqrt{t(1-e^{-2\kappa t})/2\kappa}$, $\tilde{c}(t) = c(t)^2/t$ and ${\bm \sigma}_r, {\bm \sigma}_i, {\bm \beta}, {\bm \gamma}$ denoting the volatility paremeters for the interest rate, the risky assets and the liability process, respectively. Denoting by $\Q$ the probability measure derived in \eqref{Y-prob} for $t=T$ and having in mind the ALM optimization problem stated in \eqref{mv-2} it is now clear how different beliefs or estimations for any of the model parameters involved (e.g. concerning either the processes characteristics or the financial market dependences) introduce the model uncertainty issue. In this case, different beliefs are considered with respect to the set of model parameters
\begin{eqnarray}\label{model-params}
	\begin{array}{l}
	\psi_{r} := ( R_0, \kappa, {\bm \sigma}_r^{T} )^{T} \in \R_+ \times \R_+ \times \R^n_{+}\\
	\psi_{\bm S} := ( {\bm \mu}^T, {\bm \sigma}_1^{T}, {\bm \sigma}_2^T )^T \in \R^n \times \R^{n}_+\times \R_+^{n}\\
	\psi_L := (\alpha, {\bm \beta}^T, {\bm \gamma}^T)^T \in \R \times \R^{n}_+ \times \R^m_{+}\\
	\psi_{\bm \rho} := ({\bm \rho}_W, {\bm \rho}_B) \in \R^{n\times n} \times \R^{m\times m}
	\end{array}
\end{eqnarray}
with ${\bm \rho}_W, {\bm \rho}_B$ representing the correlation structures (correlation matrices in this model setting) that are displayed by the financial market factors ${\bm W}(\cdot)$ and the insurance market factors ${\bm B}(\cdot)$, respectively. The existence of different beliefs concern\-ing the characteristics of the aforementioned model framework, results in different probability models $\Q$ describing the random behaviour of ${\bm Y}_t$ at any $t$. If $N$ different opinions concerning the model parameters \eqref{model-params} are provided, then can be equivalently represented by the prior set $\mathfrak{Q}$ including the resulting $N$ different models for the distribution of ${\bm Y}$. In this case, the interested risk manager could treat robustly the ALM problem for deriving the optimal investing portfolio through the following steps:
\begin{enumerate}
	\item The risk manager allocates her/his trust to each one of the models in prior set $\mathfrak{Q} = \{ \Q_1, ..., \Q_N\}$ by providing her/his aversion preferences through the weighting vector $\w \in \Delta^{N-1}$.
	\item The robust to model uncertainty probability measure (weighted barycenter of the prior set $\mathfrak{Q}$) $\Q_*(\w)$ is determined by applying the formulae \eqref{wloc} and \eqref{wcov} for the determination of the location and dispersion parameters.
	\item The optimal investing portfolio is calculated by applying the formula \eqref{opt-port} under the probability measure derived in Step 2.
\end{enumerate}

\section{Numerical illustration}\label{sec-5}

In this section, a simulation experiment based on the model setting discussed in Section \ref{sec-4} is performed for assessing the proposed method's capabilities in recovering the optimal investing portfolio under the model ambiguity context. A fictitious financial market is considered, consisting of a bond with stochastic instantaneous rate (modelled by an OU process of the form \eqref{r-dyn}), two risky assets (stocks following the typical GBM model \eqref{S-dyn}) while the aggregate liability process of interest evolves according to the model presented in \eqref{L-dyn}. In this model setting, the resulting random variables related to the static version of the ALM problem \eqref{mv-2} are determined in \eqref{model-rvs}. An underlying true model (unknown to the decision maker) is considered, containing two stochastic factors ${\bm W}(\cdot) = (W_1(\cdot), W_2(\cdot))$ affecting the financial market and two stochastic factors ${\bm B}(\cdot) = (B_1(\cdot), B_2(\cdot))$, independent to the financial market factors, affecting the insurance market. The initial wealth amount to be used for constructing the investing portfolio is $X_0 = 1000000.00\$$. The related true model parameters are illustrated in Table \ref{tab-1}.

\begin{table}[ht!]\small
\centering
\begin{tabular}{rrrr}
	\hline\hline
	\bf Interest Rate & \bf Risky Asset 1 & \bf Risky Asset 2 & \bf Aggregate Liability\\
	\bf (OU model)  & \bf (GBM model) & \bf (GBM model) & \bf (BM model)\\
	\hline
	$r_0 = 0.02$         & $  s_1 = 50.00$     & $s_2 = 50.00$         & $\ell_0 = 0.00$\\
	$R_0 = 0.02$         & $\mu_1 = 0.05$       & $\mu_2 = 0.10$       & $\alpha = 1000000.00$ \\
	$\kappa = 0.60$      & $\sigma_{1,1}= 0.02$ & $\sigma_{2,1}=0.05$  & $\beta_{1} = \beta_2 = 0.00$\\
	$\sigma_{r,1}= 0.005$& $\sigma_{1,2}=0.02$  & $\sigma_{2,2}=0.05$  & $\gamma_1 = \gamma_2 = 80000.00$\\
	$\sigma_{r,2}= 0.005$&  				    & 					   & \\
	\hline\hline
\end{tabular}
\caption{True model parameters used for the numerical experiments}\label{tab-1}
\end{table}

The model uncertainty framework is introduced by generating $N$ different priors perturbing the true model parameters through appropriate noise terms. In particular, perturbations from the true model parameters in the setting of \eqref{model-params} are generated through the simulation scheme
\begin{eqnarray}\label{sim-scheme}
	\begin{array}{lll}
	\psi_{r,j} = \psi_r + \eta_{r,j}, & \eta_{r,j} \sim \mathcal{U}([u_r^{(l)}, u_r^{(u)}]), & j=1,...,N \\
	\psi_{{\bm S},j} = \psi_{\bm S} + \eta_{{\bm S},j}, & \eta_{{\bm S},j} \sim \mathcal{U}([u_{\bm S}^{(l)}, u_{\bm S}^{(u)}]), & j=1,...,N \\
	\psi_{L,j} = \psi_L + \eta_{L,j}, & \eta_{L,j} \sim \mathcal{U}([u_L^{(l)}, u_L^{(u)}]), & j=1,...,N \\
	\psi_{{\bm \rho},j} = \psi_{\bm \rho} + \eta_{{\bm \rho},j}, & \eta_{{\bm \rho},j} \sim \mathcal{U}([u_{\bm \rho}^{(l)}, u_{\bm \rho}^{(u)}]), & j=1,...,N
	\end{array}
\end{eqnarray}
with the terms $\eta_{k}$ for $k = r, {\bm S}, L, {\bm \rho}$ representing appropriate uniform distributed error terms and $u_{k}^{(l)}, u_{k}^{(u)}$ denoting the related lower and upper bounds, respectively (depending on the parameter type). For each $j$ a new probability measure $\Q_j$ is created as a perturbation from the ``true" model (but unknown), i.e.
$$ \psi_j := ( \psi_{r,j}, \psi_{{\bm S}, j}, \psi_{L,j}, \psi_{{\bm \rho},j}) \mapsto \Q_j, \,\,\, j=1,...,N, $$
and all of them constitute the generated prior set $\mathfrak{Q}$. Different choices for the lower and upper bound for each parameter of interest are accordingly selected in orded to generate three different homogeneity levels for the prior set (high, medium and low) while different sizes for the priors set are also considered ($N=1, 2, 3, 5, 10, 30$). The different levels of homogeneity and prior set sizes allows for assessing the robustness of the approach in retrieving the true investing portfolio and its sensitivity with respect to the various noise levels.

The numerical experiment is performed by generating the prior set for each one of the cases under consideration, obtaining the aggregate model $\Q_*$ under equal weights and then calculating the optimal portfolio through the relation \eqref{opt-port} after generating $M=100,000$ random samples from $\Q_*$. To avoid extreme cases and test the median behaviour of the scheme, each scenario generation is repeated $K=10,000$ times and therefore the results refer to the mean estimate provided by the discussed approach (rather than to a single generation output which could be rather misleading). For the sake of convenience, $\zeta=0$ is considered, while for assessing the performance of the method at each case are calculated: the mean expected surplus amount (i.e. $\E_{\Q_*}[X_T-L_T]$), and the standard deviation of the surplus amount (i.e. $\sqrt{\E_{\Q_*}[(X_T-L_T)^2]}$) while in the parantheses are displayed the relevant standard errors of estimation with respect to the estimation under the true model parameters. The results for all cases are summarized in Table \ref{tab-2}. Moreover, to check the robustness of the estimates provided by the method as the prior set size grows, in Table \ref{tab-3} are illustrated the 95\% confidence intervals for: (a) the expected value of the surplus amount and (b) the ratio of the estimated surplus variance to the true surplus variance value.

\begin{table}[ht!]\scriptsize
	\centering
	\begin{tabular}{l|ccc|r|r}
		\hline\hline
		\bf \# of& \multicolumn{3}{c|}{\bf Wealth Allocation} & \bf Surplus Amount& \bf Surplus Amount \\
		\bf Priors & \bf Bond & \bf Stock 1 & \bf Stock 2& \bf Exp. Value (\$) & \bf Std. Dev. (\$)\\
		\hline
		\multicolumn{6}{c}{\it True Model}\\
		\hline
		- & 76.17\% & 72.44\% & -48.61\% &  -42.65\,\,\,\,\,\,\,\,\,\,\,\, (0.28) &  138585.89\,\,\,\, (10.19)\\
		\hline
		\multicolumn{6}{c}{\it High Homogeneity}\\
		\hline
            N = 1  &71.53\%   &78.25\%  &-49.78\%  &190.77\,\,\,\,\,\,\,\,\, (15.28)  &138625.36\,\,\,\, (99.46)\\
            N = 2  &73.09\%   &76.26\%  &-49.36\%  & 86.70\,\,\,\,\,\,\,\,\,\,\,\, (9.13)  &138499.23\,\,\,\, (68.57)\\
            N = 3  &75.03\%   &73.98\%  &-49.00\%  & 50.08\,\,\,\,\,\,\,\,\,\,\,\, (7.07)  &138597.02\,\,\,\, (55.85)\\
            N = 5  &76.25\%   &72.49\%  &-48.74\%  & 23.68\,\,\,\,\,\,\,\,\,\,\,\, (4.63)  &138550.81\,\,\,\, (44.54)\\
            N = 10 &76.28\%   &72.42\%  &-48.70\%  &-15.58\,\,\,\,\,\,\,\,\,\,\,\, (3.33)  &138618.56\,\,\,\, (32.24)\\
            N = 30 &75.74\%   &73.04\%  &-48.78\%  &-40.49\,\,\,\,\,\,\,\,\,\,\,\, (2.12)  &138559.28\,\,\,\, (20.23)\\
		\hline	
		\multicolumn{6}{c}{\it Medium Homogeneity}\\
		\hline
            N = 1   &65.51\%   &88.03\%  &-53.54\%  &-236.28\,\,\,\,\,\,\, (125.06) &138609.44\, (307.88)\\
            N = 2   &78.62\%   &70.43\%  &-49.06\%  & -120.35\,\,\,\,\,\,\,\,\,\, (61.58) &138630.81\, (228.14)\\
            N = 3   &74.06\%   &75.88\%  &-49.94\%  & -93.30\,\,\,\,\,\,\,\,\,\, (36.88) &138543.84\, (177.92)\\
            N = 5   &74.35\%   &75.45\%  &-49.81\%  & -69.44\,\,\,\,\,\,\,\,\,\, (26.33) &138478.29\, (139.25)\\
            N = 10  &77.94\%   &70.38\%  &-48.33\%  & -50.29\,\,\,\,\,\,\,\,\,\, (15.85) &138500.67\,\,\,\, (99.64)\\
            N = 30  &77.41\%   &70.76\%  &-48.17\%  & -45.32\,\,\,\,\,\,\,\,\,\,\,\,\, (9.88) &138409.98\,\,\,\, (55.97)\\

		\hline	
		\multicolumn{6}{c}{\it Low Homogeneity}\\
		\hline
            N = 1   &146.96\%    &-11.03\%   &-35.93\%    &-13827.14\, (13867.98)  &140337.82\, (565.53)\\
            N = 2   &83.93\%     &59.59\%    &-43.52\%    &  -638.72\,\,\,\,\,\,\, (252.74)  &138565.96\, (402.66)\\
            N = 3   &84.72\%     &60.08\%    &-44.80\%    &  -489.91\,\,\,\,\,\,\, (224.25)  &138308.13\, (319.58)\\
            N = 5   &83.78\%     &62.85\%    &-46.64\%    &  -310.36\,\,\,\,\,\,\, (141.44)  &139091.92\, (247.67)\\
            N = 10  &81.38\%     &65.42\%   &-46.81\%    &  -120.92\,\,\,\,\,\,\,\,\,\, (79.68)  &138471.82\, (170.73)\\
            N = 30  &75.94\%     &71.50\%    &-47.44\%    &  -61.63\,\,\,\,\,\,\,\,\,\, (39.14)   &138550.90\,   (102.91)\\
		\hline\hline
	\end{tabular}
	\caption{Optimal investing portfolios and the related metrics for the true model (unknown) and the ones obtained under the different homogeneity levels and prior set sizes}\label{tab-2}
\end{table}

\begin{table}[ht!]\scriptsize
\centering
\begin{tabular}{l|rr|rr}
\hline\hline
\bf \# of & \multicolumn{2}{c}{\bf Surplus Amount} & \multicolumn{2}{|c}{\bf Surplus Variance} \\
\bf Priors & \bf Lower Bound & \bf Upper Bound & \bf Lower Bound & \bf Upper Bound\\
\hline
\multicolumn{5}{c}{\it High Homogeneity}\\
\hline
N = 1     &-228.02    &1598.57    &0.959   &1.039\\
N = 2     &-145.27    & 994.16    &0.968   &1.029\\
N = 3     &-129.94    & 726.47    &0.976   &1.024\\
N = 5     &-103.13    & 442.72    &0.979   &1.019\\
N = 10    &-101.83    & 294.32    &0.986   &1.015\\
N = 30    &-100.10    & 161.11    &0.991   &1.008\\
\hline
\multicolumn{5}{c}{\it Medium Homogeneity}\\
\hline
N = 1    &-10631.53   &7492.96    &0.865   &1.138\\
N = 2    & -4208.51   &4660.98    &0.902   &1.102\\
N = 3    & -2402.89   &2535.51    &0.922   &1.078\\
N = 5    & -1593.16   &1894.37    &0.936   &1.061\\
N = 10   &  -972.31   &1137.25    &0.956   &1.045\\
N = 30   &  -612.19   & 563.08    &0.974   &1.025\\
\hline
\multicolumn{5}{c}{\it Low Homogeneity}\\
\hline
N = 1    &-30265.44   &30399.51   &0.772   &1.277\\
N = 2    &-18855.45   &16866.59   &0.829   &1.199\\
N = 3    &-15045.58   &13658.66   &0.859   &1.144\\
N = 5    &-10463.15   & 9977.13   &0.885   &1.109\\
N = 10   & -6401.53   & 4533.95   &0.923   &1.070\\
N = 30   & -3240.59   & 1994.64   &0.954   &1.046\\
\hline\hline
\end{tabular}
\caption{Lower bound (2.5\% percentile) and upper bound (97.5\% percentile) values for the surplus amount (left) and for the ratio of the estimated surplus variance to the actual (right)}\label{tab-3}
\end{table}

Comparing the results to the optimal solutions obtained under the true model, it seems that the applied method displays a remarkable performance in retrieving the true optimal investing portfolios and the surplus characteristics up to the second moment, even in the low homogeneity case of the provided information set. In all cases, the standard errors seems to be quite low for $N>1$ and to display quite acceptable values among the various information homogeneity levels, indicating a robust behaviour of the method with respect to the information heterogeneity. When smaller prior set sizes are used, the standard errors seems to naturally increase due to the effect of possible extreme opinions, while the wealth proportion allocations significantly differ from the true ones, especially in cases of one prior opinions in the medium and low homogeneity scenarios. For prior sets constituting of five or more models, the method's accuracy seems to be significantly improved, especially in recovering the true optimal wealth amount allocations to the various hedging instruments. The successful handling of the multiple information from the proposed scheme becomes more clear in Table \ref{tab-3} where as the prior set size grows, the gap in the expected surplus amount and the surplus variance optimal estimations becomes smaller. This is an evidence of a robust behaviour indicating that the more the available models, the more robust the derived optimal decision becomes to model ambiguity. This is much more clearer if one compares the results of each one of $N>1$ priors cases to the single prior case ($N=1$). The single prior case also represents the situation that one fully allocates her/his trust to a single model from the prior set, increasing the financial decision's vulnerability to the prior model inefficiencies.

In general, when the prior set provides a sufficient amount of information around the true model, the discussed method seems capable of successfully filtering out the information noise and providing estimates quite close to the true ones. However, since the results significantly depend on the quality of the prior models provided, the interested decision maker should be aware of the possibility that the whole prior set to provide quite misleading models. In such a case, the pure barycenter as employed here as an aggregate tool may fail and aversion preferences with respect to the whole prior set might be introduced in the sense discussed in \cite{papayiannis2018convex}, and then obtaining worst-case investment strategies. However, when the prior set is quite representative of the situation but some models may be less accurate, the proposed method 's accuracy can be further improved by accompanying the approach with proper weight selection principles for the barycenter model (see e.g. \cite{papayiannis2018learning, papayiannis2018model}) as more data becomes available which can be used to assess the validity of the models.

\section{Conclusions}

In this paper, the asset liability management problem has been revisited under the framework of model ambiguity when a multitude of prior models are available. A working framework for treating robustly the model uncertainty issues has been presented, under which the related optimal investing portfolio is derived. The aggregation of the information provided by the available prior models concerning the discounting measure is the key intermediate step which relies on the notion of the Wasserstein barycenter. Results in semi-closed form and characterization of the optimal and robust to model ambiguity portfolio for standard modelling approaches used in the ALM problem are provided. Finally, the proposed method's performance has been assessed in a numerical experiment with various noise levels considered for the prior models with very satisfactory results.

\section*{Acknowledgments}
The author would like to thank the editor and the associated reviewers for their constructive comments and suggestions that helped in improving the quality of this manuscript.


\end{document}